\documentclass{article}
\usepackage{ijcai25}

\usepackage{times}
\usepackage{soul}
\usepackage{url}
\usepackage[hidelinks]{hyperref}
\usepackage[utf8]{inputenc}
\usepackage[small]{caption}
\usepackage{graphicx}
\usepackage{amsmath}
\usepackage{amsthm}
\usepackage{booktabs}
\usepackage{algorithm}
\usepackage{algorithmic}
\usepackage[switch]{lineno}

\usepackage{amssymb}
\usepackage[T1]{fontenc}
\usepackage{subcaption}

\usepackage{float}

\renewcommand\vec{\mathbf}

\newcommand{\calN}{\mathcal{N}}
\newcommand{\calS}{\mathcal{S}}
\newcommand{\calA}{\mathcal{A}}
\newcommand{\calO}{\mathcal{O}}
\newcommand{\calT}{\mathcal{T}}
\newcommand{\calR}{\mathcal{R}}

\newcommand{\calD}{\mathcal{D}}
\newcommand{\E}{\mathbb{E}}
\newcommand{\R}{\mathbb{R}}

\title{Controllable Complementarity: Subjective Preferences in Human-AI Collaboration}

\author{
Chase McDonald
\And
Cleotilde Gonzalez
\affiliations
Carnegie Mellon University\\
\emails
\{chasemcd, coty\}@cmu.edu
}

\begin{document}

\maketitle

\begin{abstract}
    Research on human-AI collaboration often prioritizes objective performance. However, understanding human subjective preferences is essential to improving human-AI complementarity and human experiences. We investigate human preferences for controllability in a shared workspace task with AI partners using Behavior Shaping (BS), a reinforcement learning algorithm that allows humans explicit control over AI behavior. 
    In one experiment, we validate the robustness of BS in producing effective AI policies relative to self-play policies, when controls are hidden. In another experiment, we enable human control, showing that participants perceive AI partners as more effective and enjoyable when they can directly dictate AI behavior. Our findings highlight the need to design AI that prioritizes both task performance and subjective human preferences. By aligning AI behavior with human preferences, we demonstrate how human-AI complementarity can extend beyond objective outcomes to include subjective preferences.\footnote{Experiment ethical approval was received by the university's Institutional Review Board.}
\end{abstract}

\section{Introduction}

The past several decades have seen a significant increase in research focusing on human-AI teams (HATs) and human-AI collaboration. Both refer to situations where there exists at least one human and an autonomous agent, where autonomous agents play a significant role and are treated as teammates \cite{o2022human}. Conceptually, HATs have existed for several decades; since computers became commonplace, researchers have been interested in how they can be used to work alongside humans \cite{nass1996can}. Indeed, as AI systems have advanced, it has become increasingly clear that AI has usage beyond just that of a tool to support human activities; it can serve the role of a teammate.


The focus in the HAT literature is typically on improving team performance, often through human-AI complementarity, where the combination of humans and AI exceeds the performance or capabilities of either in isolation \cite{bansal_does_2021,steyvers2023three}. The emphasis in this definition is on objective performance. However, this neglects other considerations, particularly those related to subjective human preferences. The stance that we take is that the scope of objective performance in human-AI interaction is too narrow and that it can be expanded to account for subjective preferences: human-AI complementarity emerges when subjective preferences are met alongside objective performance, enhancing human experiences beyond what would be possible without AI.

It may not always be the case that humans seek only to reach some final outcome, but also seek a particular path to that outcome---the experience along the way is important \cite{gandhi2024beliefs}. Prior work has pointed out the importance of accommodating human preferences and capabilities in building HATs \cite{zhao2023learning,van2019six}, although there is a gap in the literature on the extent to which humans are will accept degradations in objective outcomes to improve their subjective utility. We can also consider this from an alternative perspective. Given a fixed objective level of performance, what factors contribute to the human \emph{subjective preferences} of their HAT experience? How important are these relative to their objective performance? As AI technologies progress, it is not difficult to imagine scenarios in which HATs are worse than teams composed only of AI agents; in many settings this is already the case, with AI teams reaching performance that exceeds human performance in a number of tasks \cite{silver2017mastering,brown2018superhuman,berner2019dota,vinyals2019grandmaster}. The incorporation of humans, particularly in low-stakes settings, may then be to satisfy some subjective preferences, rather than on optimal objective outcomes. 


Our primary focus in this work is on subjective preferences for the control of AI partners. Control has been reported to be a desired component of AI teammates \cite{klien2004ten,lim2009assessing,myers2001human}. Although highlighted as an important component, control has not played a significant role in the existing HAT literature \cite{tsamados2024human} and the need for additional methods to enable control has been noted \cite{teaming2022state}.

Our aim is to address open questions in human-AI teaming by exploring the following research questions:
\begin{itemize} \item \textbf{RQ1}: In shared workspace tasks, do humans have preferences for the controllability of their AI teammates?
\item \textbf{RQ2}: How do humans balance subjective preferences and task performance? Furthermore, how can we account for subjective preferences to enhance human experiences in human-AI collaboration?
\end{itemize}

We contribute to the machine learning literature by developing \emph{controllable} AI policies to investigate these questions. Specifically, we use reinforcement learning \cite{sutton_reinforcement_2018} as our baseline and augment the training regime to allow explicit human control over the learned policy through a method we refer to as Behavior Shaping (BS). Through a user study, we demonstrate the efficacy of BS in creating robust AI partners capable of collaborating effectively with humans, offering an improvement relative to the standard self-play baseline.

Then, in a second study, we use BS to address the outlined research questions. By systematically manipulating participants’ ability to use and observe controls over AI teammate behavior, we examine how these manipulations influence subjective evaluations of the AI partner.
\section{Related Work}

We first outline existing work relevant to the human-AI teaming literature and the development of collaborative AI.

\paragraph{Human-AI Teaming.}

There has been significant effort to outline frameworks and desiderata for the characteristics of HATs, which is often based on the literature on human-human teams or human preferences in surveys \cite{klien2004ten,hauptman2023adapt,endsley2017here,mcneese2021my}. A noteable amount of effort has also investigated appropriate or desired levels of autonomy for AI teammates \cite{schermerhorn2009dynamic,ball2012explorations,alan2014field,endsley2017here,sundar2020rise,salikutluk2024evaluation}. Although there is a significant amount of research in the AI literature on how to build effective and adaptive AI teammates \cite{carroll2019utility,strouse2021collaborating,zhao2023learning,zhang2022you,yu2023learning}, there are limited empirical evaluations and validations of the aforementioned frameworks and desiderata in shared workspace settings \cite{salikutluk2024evaluation}, where humans and AI act independently in an environment together. Therefore, further investigation and evaluation of human behavior and preferences is needed in the context of HATs.

A commonly reported attribute of AI teammates in HATs is that they must be \emph{predictable}, particularly when they have high levels of autonomy \cite{hauptman2023adapt}. Predictability arises from the need to coordinate behavior in collaborative settings and is critical in teams with any type of composition. In fact, predictability is closely related to theory of mind, in which teammates must be able to anticipate others' behaviors to plan their actions accordingly \cite{klien2004ten,byom2013theory,dechurch2010cognitive}. Being able to anticipate what a teammate will do and take actions that complement anticipated behavior is critical for team performance. Existing work has demonstrated that developing models of humans and using them to allow AI agents to predict human behavior has increased team performance \cite{wu2021too,westby2023collective}. Little empirical work has investigated how the predictability of agents impacts objective and subjective outcomes for HATs, as well as on human preferences for predictability beyond survey responses \cite{o2022human,zhang2021ideal}.

\paragraph{Reinforcement Learning \& Zero-Shot Coordination}


In collaborative domains, a desirable attribute of AI partners is the ability to be robust and adaptive to the behaviors and preferences of humans \cite{strouse2021collaborating}. An AI partner or teammate should maintain its effectiveness regardless of who or what it is interacting with and be able to complement the skills and preferences of those it interacts with. This is typically tested through zero-shot coordination \cite{hu_other-play_2021}, which evaluates the degree to which a strategy learned in isolation is effective with an unseen partner. 

Diversity in training experience has been shown to improve the generalization abilities \cite{strouse2021collaborating,lupu2021trajectory} of AI to diverse populations of humans. This is particularly important in human-AI interaction, as we would like to develop efficient and robust regimes that allow us to train policies that are effective at generalizing to human partners. Several approaches have been proposed that tackle the problem in collaborative domains, with the most successful typically based off the central notion that the policy should be trained with a sufficiently diverse set of training partners. To this end, \cite{strouse2021collaborating} proposed using a population of policies using different network architectures, random seeds, and levels of experience as a training population. 


Additional work based on this approach has structured the construction of that population in more nuanced ways \cite{lupu2021trajectory,zhao2023maximum,zhao2023learning,yu2023learning,charakorn2022generating}. Our proposed approach implicitly captures the characteristics of a population that each of the past works aims for. With a combination of sufficiently expressive behaviors and sampling schemes for the \emph{single} policy we train, it is possible to achieve sufficient behavioral diversity. The most closely related approach is Any-Play \cite{lucas2022any}, which similarly trains a conditional policy, but with the goal of maximimally discernable behavioral expressions given the latent variable on which it is conditioned. The output of Any-Play is an ``accommodator'' agent that learns to collaborate with the conditional policy, whereas the output of BS is the conditional policy. 



\section{Methods}

\newcommand{\figheight}{1.7cm} 

\begin{figure*}
    \centering
    \begin{subfigure}{0.19\linewidth}
        \centering
        \includegraphics[height=1.7cm]{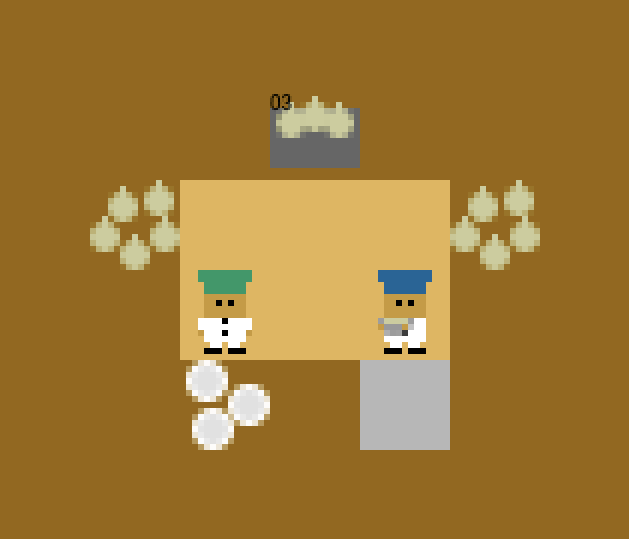}
        \caption{Cramped Room}
    \end{subfigure}%
    \begin{subfigure}{0.19\linewidth}
        \centering
        \includegraphics[height=\figheight]{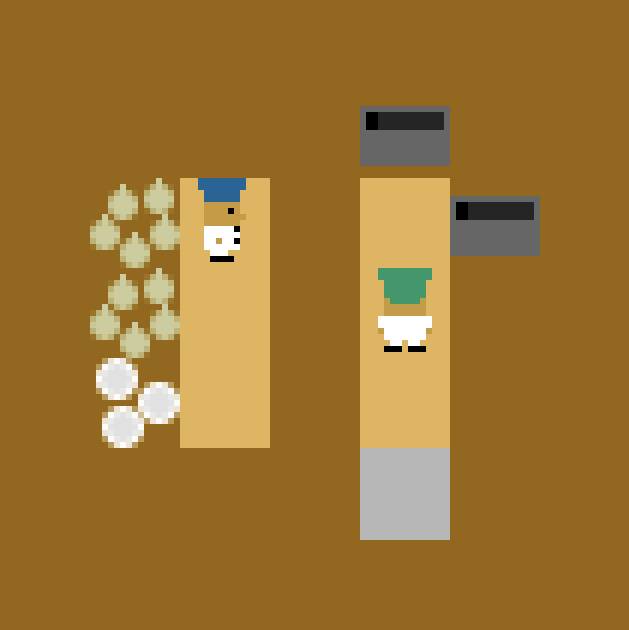}
        \caption{Forced Coordination}
    \end{subfigure}%
    \begin{subfigure}{0.19\linewidth}
        \centering
        \includegraphics[height=\figheight]{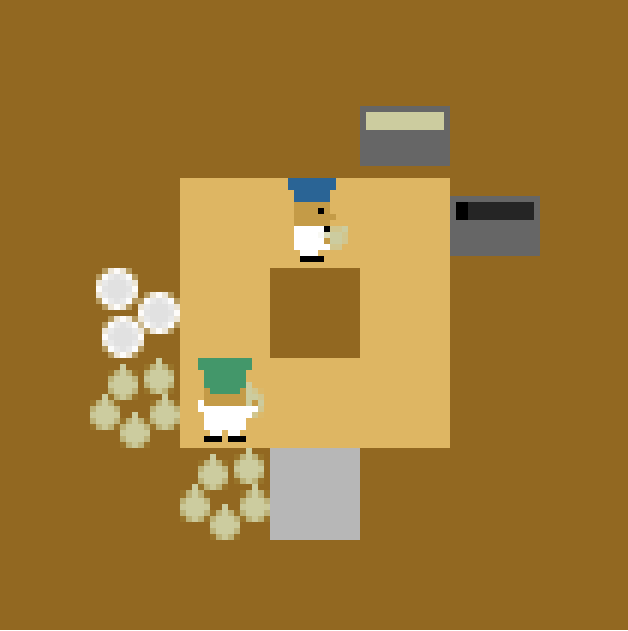}
        \caption{Coordination Ring}
    \end{subfigure}%
    \begin{subfigure}{0.19\linewidth}
        \centering
        \includegraphics[height=\figheight]{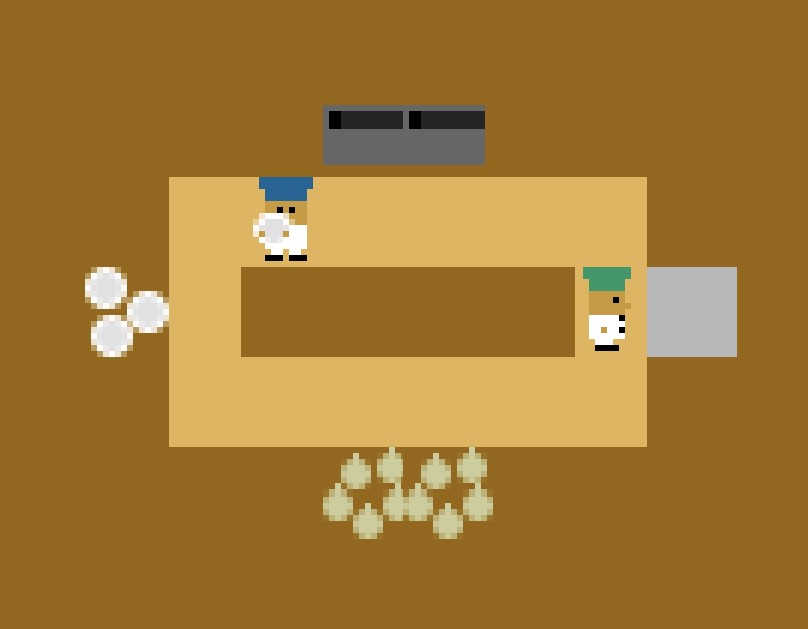}
        \caption{Counter Circuit}
    \end{subfigure}%
    \begin{subfigure}{0.2\linewidth}
        \centering
        \includegraphics[height=\figheight]{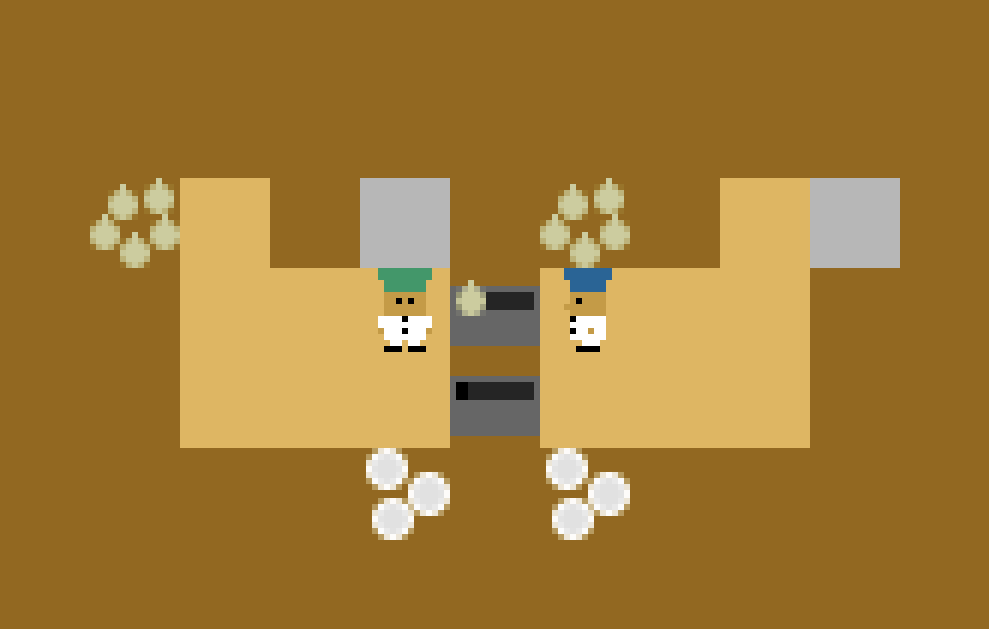}
        \caption{Asymmetric Advantages}
    \end{subfigure}

    \caption{The five overcooked layouts. In our experiments, the AI always takes the role of the chef in the green hat.}
    
    \label{fig:overcooked_layouts}

\end{figure*}

\subsection{Preliminaries}

Environments where agents (both human and autonomous) observe information about the state of the world, take actions, and observe outcomes can be characterized as decentralized (partially observable) Markov decision processes (Dec-POMDPs). With $N$ agents, each POMDP is described by a tuple $\langle \calN, \calS,  \{\calA^{i}\}_{i\in\calN}, \calO, \calT, \{\calR^{i}\}_{i\in\calN}, \{\Omega^{i}\}_{i\in\calN}, \{\gamma^{i}\}_{i\in\calN} \rangle$, where $\calN=\{1,...,N\}$ is the set of $N>1$ agents, $\calS$ is the state space, $\calA^{i}$ the action space for agent $i$, $\calO$ the observation space, $\calT:\calS\times\calA\to\calS$ the transition function, $\calR^{i}:\calS\times\calA\to\R$ the reward function for agent $i$, $\Omega^{i}:\calS\to\calO$ the observation function for agent $i$ and $\gamma^i\in\R$ for discount factor for agent $i$. By $\calA$, $\calR$, $\Omega$, and $\gamma$ we represent the joint action spaces, the reward functions, the observation functions, and the discount factor for all agents $i\in1,...,N$.

At a particular time step $t\in \{1,...,T\}$, we have the state described by $s_t\in\mathcal{S}$. The set of observation functions map the state to a set of agent observations $\Omega\left( s_t \right)=\vec{o}_t\in\calO$. Given the observations $\vec{o}_t$, each agent selects an action $a^i_t\in\calA^i$. The transition function then determines the subsequent state through $\calT(s_t, \vec{a}_t)=s_{t+1}$. Similarly, the reward for each agent $i$ is determined by its corresponding reward function $\calR^i(s_t, \vec{a}_t)=r^i_t\in\R$. This process then repeats until the final step $T$ is reached. A trajectory $\tau=(\vec{s}_0, \vec{a}_0, \vec{s}_1, \vec{a}_1, ...)$ then denotes a sequence of states and actions that fully describes the experienced state transitions and actions in an episode of interaction.

In a Dec-POMDP, each agent learns a behavior policy $\pi^{i}:\calO\to P(\calA^{i})$ that maps observations to action probabilities: $\pi^{i}(a|o)$ depicts the probability of agent $i$ selection action $a\in\calA^{i}$ after observing $o\in\calO$. Each policy $\pi^i$ is learned to maximize the performance measure shown in Equation~\ref{eqn:multi_agent_objective}, where $\pi^{-i}$ denotes the policies of all agents other than $i$.

\begin{equation} \label{eqn:multi_agent_objective}
    \max_{\pi^{i}} \E_{\tau\sim \langle\pi^{i}, \pi^{-i}\rangle}\left[ \sum_{t>0} \left(\gamma^{i}\right)^t \mathcal{R}^{i}(s_t, \vec{a}_t) \right]
\end{equation}

In reinforcement learning (RL), each policy $\pi$ is optimized for the objective in Equation~\ref{eqn:multi_agent_objective}. Across experiments, we use Proximal Polixy Optimization \cite{schulman_proximal_2017} as our training algorithm. As our baseline, we utilize self-play (SP) training. This is a paradigm where all agents share the same set of parameters---the policy learns to take actions from the perspective of all agents.

\subsection{Behavior Shaping (BS)}

Typically, RL policies select actions to maximize a static reward function given an observation of the world. In BS, a policy is conditioned not only on an observation of the environment but also on a set of weights. Each of these weights correlates with additional reward functions that represent desired (or undesired) behaviors. If a policy learns to maximize its reward conditioned on different values of these weights, it will learn different behaviors when the weights are manipulated. If the reward functions are human interpretable, we have the ability to manually alter the weights that the agent observes to intentionally change its behavior. 

With BS, we augment both observation and reward with a behavioral target. Rather than observing only $o^i_t$, policies are also conditioned on values that represent desired behaviors. We let $\Psi$ be the space of behavioral reward functions. For each behavioral reward function $\psi\in\Psi^i$, there is a corresponding weight $\omega\in\R$, sampled from a distribution $\calD_\psi$ at the beginning of each episode. The policy $\pi^i$ and the behavioral reward functions $\psi$ are then conditioned on this weight. With $K$ behavioral reward functions, action selection becomes $\pi^i(a_t|o^i_t,\vec{\omega}^i)$ and the reward at each step includes contributions from the behavioral reward functions: $r^{'i}_t=\calR^i(s_t, \vec{a}_t) + \sum_k \psi^i_k(s_t, \vec{a}_t | \omega^i_k)$. 

The objective of each agent $i$ is then augmented to account for these additional rewards and the sampling of $\vec{\omega}^i$. The procedure used in training policies with BS is shown in Algorithm~\ref{alg:BS}, where $\psi^i(s_t, \vec{a}_t|\vec{\omega}^i)=\sum_{k\in \{1,...,K\}} \psi^i_k (s_t, \vec{a}_t | \omega^i_k)$. \\

This formalization aligns directly with the notion of a contextual MDP \cite{hallak2015contextual}, where an MDP is augmented with a context that alters both the observation and reward function. BS can be thought of as inducing a contextual MDP where the combination of $\omega$ values represents a particular context. The particular requirement of BS is that each individual $\omega$ is interpretable in its behavioral implications to allow for behavioral control by humans.



\begin{algorithm}[H]
\caption{BS Training Procedure}
\label{alg:BS}
\begin{algorithmic}[1]
\STATE Initialize behavioral reward functions $\psi^i \forall i \in \mathcal{N}$
\STATE Initialize weight distributions $\mathcal{D}_{\psi^i} \forall i \in \mathcal{N}$
\FOR{each episode}
    \STATE Initialize $s_0$
    \STATE $\omega^i \sim \mathcal{D}_{\psi^i} \forall i \in \mathcal{N}$
    \FOR{$t \in \{0, \ldots, T\}$}
        \STATE $o^i_t \gets \Omega^i(s_t) \forall i \in \mathcal{N}$
        \STATE $a^i_t \sim \pi^i(a \mid o^i_t, \omega^i) \forall i \in \mathcal{N}$
        \STATE $r^i_t \leftarrow \mathcal{R}^i(s_t, \vec{a}_t) + \psi^i(s_t, \vec{a}_t \mid \omega^i) \forall i \in \mathcal{N}$
        \STATE $s_{t+1} \sim \mathcal{T}(s_t, \vec{a}_t)$
    \ENDFOR
\ENDFOR
\end{algorithmic}
\end{algorithm}

We achieve the two goals of (1) a robust and adaptive policy and (2) policy controllability through the random sampling of $\vec{\omega}$ during training. Both arise from the fact that we represent an arbitrarily diverse population of agents in a single conditional policy. By randomly sampling the behavioral weights $\omega$ on which the policy is conditioned, a wide variety of behaviors are encouraged that simulate training with a diverse population of agents. Controllability also arises from the single policy because at test time, the behavioral weights can be set manually to elicit desired behavioral patterns.

There are several important dimensions that will impact the diversity of behavior that a policy is capable of and the level of control over a policy. Specifically, both $K$ and $\calD_\psi$ regulate the breadth of behavioral diversity possible in a policy. The number of behavioral dimensions defined by $K$, presuming that each dimension represents a sufficiently distinct behavior, directly relates to the number of possible different behavioral expressions. Similarly, $\mathcal{D}_\psi$ defines the breadth of expression within a single behavior. 

Importantly, a fundamental component of BS is the human interpretability of $\psi$ and $\mathcal{D}_\psi$. Methods for skill and behavior discovery exist (e.g., \cite{yang2024hierarchical}); however, they may not align with human expectations. For this reason, it is necessary to specify the desired behavioral dimensions a priori. It should be noted that existing and concurrent work has made an effort to align automated skill discovery with human preferences (e.g., \cite{hussonnois2025human}) in favor of hand-crafted reward functions due to potential brittleness. Our approach to BS makes the first step in defining a signal that can be manipulated in a way that yields expected behavior, which could become increasingly complex with more sophisticated approaches to reward signal design, such as through language models \cite{kwon2023reward}.

BS presents several benefits over past methods for building robust, collaborative agents. As previously described, the policy is expressive and the training regime facilitates expressivity and controllability in policy behavior, the latter of which is not captured explicitly by any of the previous works in the domain of human-AI collaboration. Furthermore, BS utilizes a single conditional policy, contrasting the population of policies required by most past work.

\subsection{Overcooked}

We conducted our experiments in the Overcooked-AI environment \cite{carroll2019utility}, reimplemented in CoGrid \cite{mcdonald2024cogrid}. This is a simplified version of the Overcooked video game in which players control chefs who are working to cook and deliver dishes to waiting patrons. Two chefs cook a soup by placing three onions in a pot, letting it cook, putting it in a dish, and then delivering it to the delivery zone. In this fully cooperative environment, both agents earn a reward of 1.0 when a dish is delivered. The task is designed such that effective collaboration is necessary to achieve high scores.


We use the five layouts of the environment as originally proposed by \cite{carroll2019utility} and shown in Figure~\ref{fig:overcooked_layouts}. Full details on the layouts and their characteristics are provided by \cite{carroll2019utility}. We make minor augmentations that increase episode length to 1,000 steps. 
To train all policies in Overcooked, we perform a hyperparameter search for the SP policies across environments. These parameters were fixed for BS. Training was carried out over 500 million environment steps with the best checkpoint of each, across five random seeds, being selected for inclusion in the experiments.

\subsection{BS in Overcooked}

To implement BS in Overcooked, we condition the policies on the weights $\omega$ of three simple behavioral reward functions. The first, $\psi_1$, is the delivery act reward. While always receiving the extrinsic delivery reward dictated by the environment, we define a similar reward that is not shared across agents: $\psi_1$ gives a reward of $\omega_1^i$ to agent $i$ if agent $i$ delivers a dish. Second, for $\psi_2$, we have the onion-in-pot reward, which gives a reward of $\omega_2^i$ each time agent $i$ places an onion in the pot. Lastly, the final behavioral reward function $\psi_3$ provides $\omega_3^i$ reward to agent $i$ if they plate a dish. Each of the weights is sampled according to $\omega\sim\calN(0, 1)$. Again, these behavioral reward functions are distinct from the extrinsic environment reward: regardless of the $\omega$ values, all agents earn a reward of 1.0 when a dish is delivered. In our human experiments, we discretize the $\omega$ values and label $\omega=-1$ as ``Discourage,'' $\omega=0$ as ``Neutral,'' and $\omega=1$ as ``Encourage.'' 
\section{Experiment 1: Zero-Shot Coordination}

To evaluate the robustness and collaborative capacity of BS agents in Overcooked, we conduct an experiment with human participants. The goal of this experiment is to test, relative to standard methods, if BS can produce a robust policy capable of human-AI interaction. This sets an important precedent for the use of these policies to study the interaction and collaboration between humans and AI. As in prior work (e.g., \cite{strouse_collaborating_2021}), evaluation of robustness and collaborative capability is carried out on the basis of objective score and subjective preferences. 

\paragraph{Design.}
Participants were recruited to play the Overcooked game with an AI partner. We recruited 59 participants (23.7\% female, 0\% other; avg. age $35.00\pm 7.21$) through Amazon Mechanical Turk. In the experiment, the possible AI partners were a standard self-play (SP) policy and a BS policy. In training, BS weights were sampled, but they were set to $\omega=0$ in the human experiment. 

In our within-subjects design, participants completed 10 pairwise comparisons between a standard self-play (SP) policy and the BS policy, two in each of the Overcooked layouts. After reading the instructions and completing a brief tutorial, participants played a 45-second round of the game with one partner, then a second round on the same layout with the other partner. They were then asked if they preferred their first or second partner. The order of the layouts and the partners in each layout were randomly assigned to the participants. We focus on a single configuration, where the human always controls the chef in the blue hat. This means that the human is always on the right in Asymmetric Advantages and left in Forced Coordination. The experiment was designed with Interactive Gym \cite{mcdonald2024cogrid}.

\paragraph{Hypotheses.} We have two primary hypotheses. The first (H1.1) is that the objective performance in human-AI pairs with BS will exceed that of pairs with SP agents. This is motivated by the fact that BS simulates a diverse population in training, making the agents more robust to unseen partners. The second (H1.2) is that BS will be subjectively preferred by human participants.

\paragraph{Results.}

\begin{figure}
    \centering
    \includegraphics[width=1.0\linewidth]{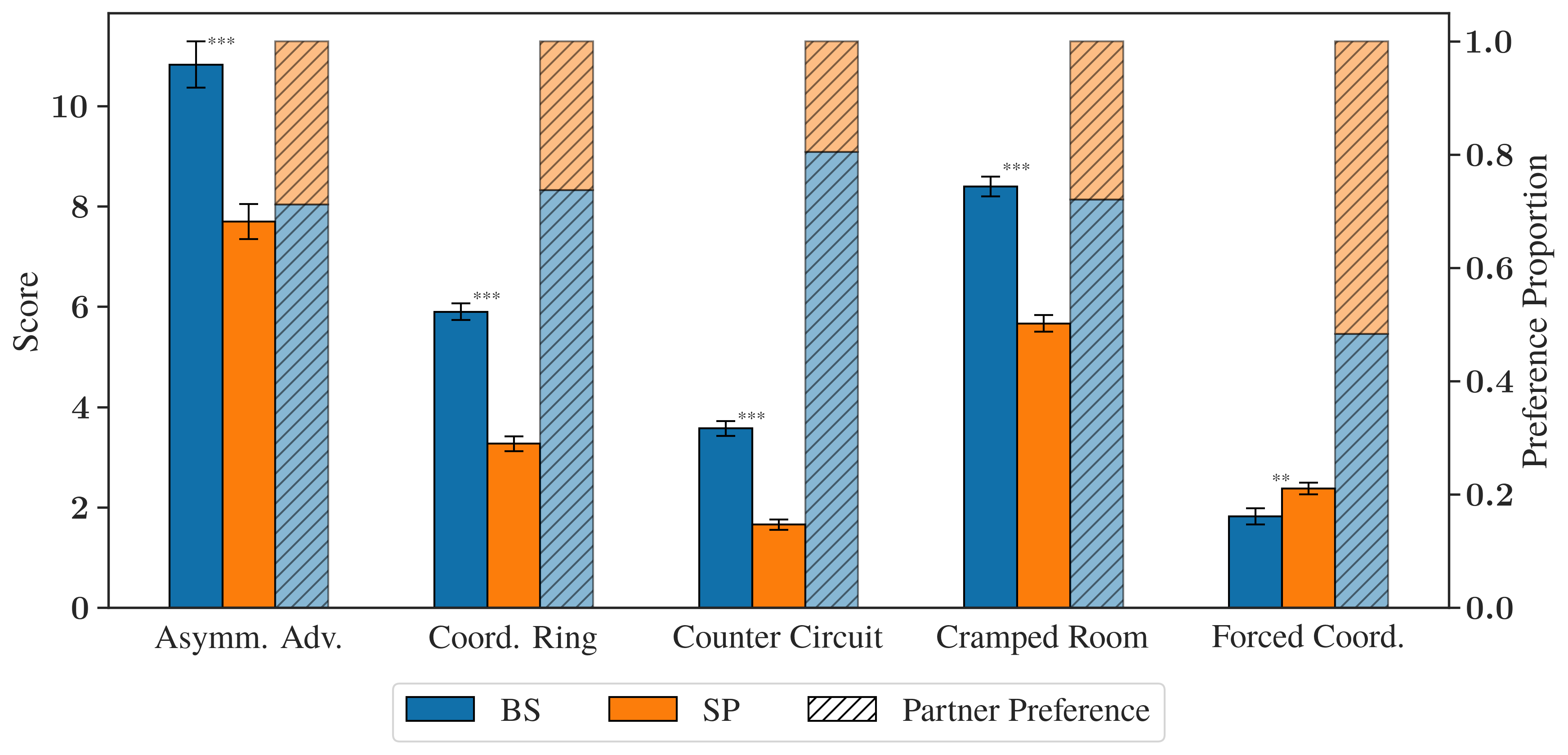}
    \caption{Average score and preference proportion by layout and AI partner type. Score by partner corresponds to the left axis. The  preference between partners corresponds to the right axis. Bars indicate standard error and stars indicate significance levels in the score differences between partner types: $***, p<0.001$ and $**, p<0.01$.}
    \label{fig:zsc_scores_preferences}
\end{figure}

The primary results in terms of objective performance are shown in Figure~\ref{fig:zsc_scores_preferences} in the two bars on the left. We find support for H1.1 on all but one layout. On all layouts, except for Forced Coordination, the score (number of dishes delivered) is significantly higher for human-BS teams compared to human-SP teams. The lack of improvement in Forced Coordination is not entirely surprising, the BS mechanisms that we randomize over only affect the chef on the right-hand side. Indeed, putting onions in the pot, plating plates, and delivering plates are all done by a single chef, so manipulating them does not encourage diverse behavior between partners. In fact, the chef on the left-hand side will have less experience to learn an effective strategy (providing supplies to the agent on the right), because its partner may sometimes have settings that do not provide a positive signal on those behaviors (e.g., discouraged from putting onions in the pot or delivering dishes). Nevertheless, we see support for H1.1 and robustness across the remaining layouts: BS was able to simulate diversity that resulted in more effective partners, relative to SP. The results in Forced Coordination underscore the importance of  appropriate behavioral reward selection.


The results in terms of subjective preferences are also shown in Figure~\ref{fig:zsc_scores_preferences} on the right axis. Aligning with objective performance, we observe subjective preferences for BS over SP in all layouts except for Forced Coordination---for the same reasons as above. After each interaction, most of the participants indicated that they preferred the BS partner over SP. This supports H1.2 in all layouts, except one. 


This experiment provides motivating results for using BS to study human-AI interaction: We demonstrate in all but one layout that both objective performance and subjective preferences favor BS over standard self-play agents. The results of this first experiment raise an important question. In line with previous work (e.g., \cite{strouse_collaborating_2021}), we ask participants about their subjective preferences. These preferences appear to align directly with objective task performance. Howveer, this is only a single factor that can influence the subjective human preferences of their AI teammate, which motivates the remainder of our investigation. What other factors influence subjective evaluations and how can we manipulate them to improve human experiences?

\section{Experiment 2: Preferences for Control} \label{sec:control_experiment}

\begin{table*}[ht]
    \centering
    {
\def\sym#1{\ifmmode^{#1}\else\(^{#1}\)\fi}
\begin{tabular}{l*{6}{c}}
\hline\hline
                    &\multicolumn{3}{c}{All Conditions}                               &\multicolumn{3}{c}{Observable $\omega$}                             \\\cmidrule(lr){2-4}\cmidrule(lr){5-7}
                    &\multicolumn{1}{c}{(1)}&\multicolumn{1}{c}{(2)}&\multicolumn{1}{c}{(3)}&\multicolumn{1}{c}{(4)}&\multicolumn{1}{c}{(5)}&\multicolumn{1}{c}{(6)}\\
                    &\multicolumn{1}{c}{Predictable}&\multicolumn{1}{c}{Effective}&\multicolumn{1}{c}{Enjoyable}&\multicolumn{1}{c}{Predictable}&\multicolumn{1}{c}{Effective}&\multicolumn{1}{c}{Enjoyable}\\
\hline
                &                     &                     &                     &                     &                     &                     \\
Fixed               &        1.76\sym{***}&        0.39         &        0.18         &                 &                 &             \\
                    &      (0.25)         &      (0.24)         &      (0.23)         &                  &                  &                  \\
\\[-0.3em]
Controllable        &        2.09\sym{***}&        1.55\sym{***}&        1.39\sym{***}&        0.01         &        0.94\sym{***}&        0.83\sym{***}\\
                    &      (0.27)         &      (0.25)         &      (0.25)         &      (0.20)         &      (0.22)         &      (0.21)         \\
\\[-0.3em]
Score               &        0.90\sym{***}&        1.25\sym{***}&        1.23\sym{***}&        0.23\sym{***}&        0.64\sym{***}&        0.63\sym{***}\\
                    &      (0.05)         &      (0.05)         &      (0.04)         &      (0.03)         &      (0.03)         &      (0.03)         \\
\\[-0.3em]
Followed Settings   &                     &                     &                     &        0.81\sym{***}&        0.55\sym{***}&        0.57\sym{***}\\
                    &                     &                     &                     &      (0.03)         &      (0.03)         &      (0.03)         \\
\\[-0.3em]
Controllable $\times$ Followed Settings&                     &                     &                     &       -0.02         &        0.18\sym{***}&        0.16\sym{***}\\
                    &                     &                     &                     &      (0.04)         &      (0.04)         &      (0.04)         \\
\hline
Observations        &       2,083         &       2,083         &       2,083         &       1,388         &       1,388         &       1,388         \\
Groups (Participants)&         116         &         116         &         116         &         116         &         116         &         116         \\
\hline\hline
\multicolumn{7}{l}{\footnotesize Standard errors in parentheses}\\
\multicolumn{7}{l}{\footnotesize \sym{*} \(p<0.05\), \sym{**} \(p<0.01\), \sym{***} \(p<0.001\)}\\
\end{tabular}
}

    \caption{Regression results for modeling participant subjective responses. Each is a mixed-effects model that also includes controls for the interaction of the $\omega$ settings, environment layout, as well as random intercepts at the participant level. In columns (1)-(3), the omitted condition is the Hidden-settings conditions and in columns (4)-(6) the Fixed condition is omitted. All survey response variables (Predictable, Effective, Enjoyable, Followed Seetings) are mean-centered.}
    \label{tab:reg_results}
\end{table*}

\begin{figure}
    \centering
    \includegraphics[width=0.9\linewidth]{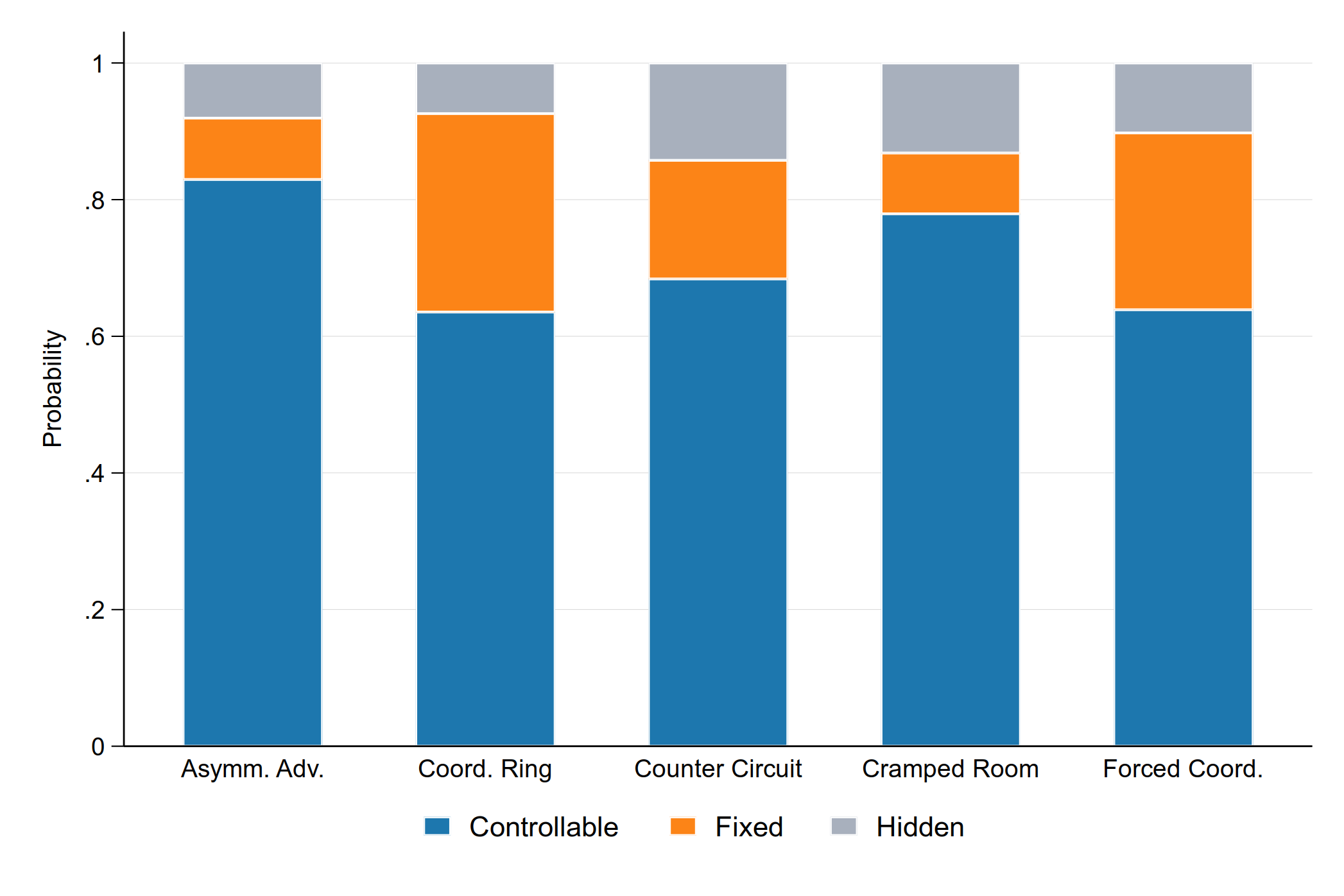}
    \caption{The predicted probability of choosing to interact with a controllable, fixed, or hidden-setting partner. The probabilities were derived from a multionomial logistic regression, which controlled for layout and all fixed behavior setting interactions, and scores with each type of partner. Across all layouts, the majority of choices are predicted to be for the controllable partner.}
    \label{fig:choice_probs}
\end{figure}

\begin{figure*}
    \centering
    \begin{subfigure}{0.4\linewidth}
        \centering
        \includegraphics[width=\textwidth]{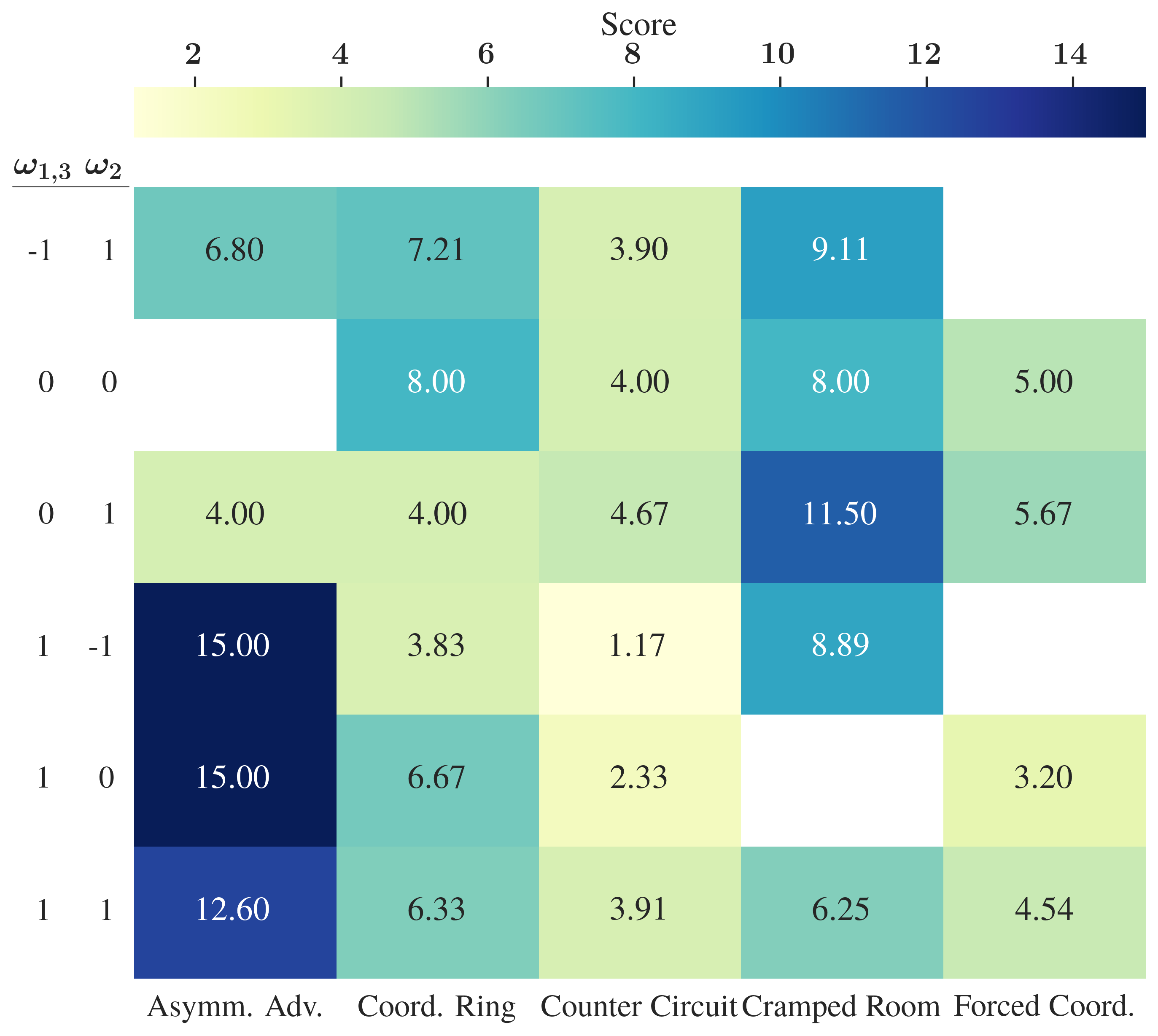}
        \caption{Average score by layout and $\omega$ values.}
        \label{fig:score_heatmap}
    \end{subfigure} 
    \begin{subfigure}{0.4\linewidth}
        \centering
        \includegraphics[width=\textwidth]{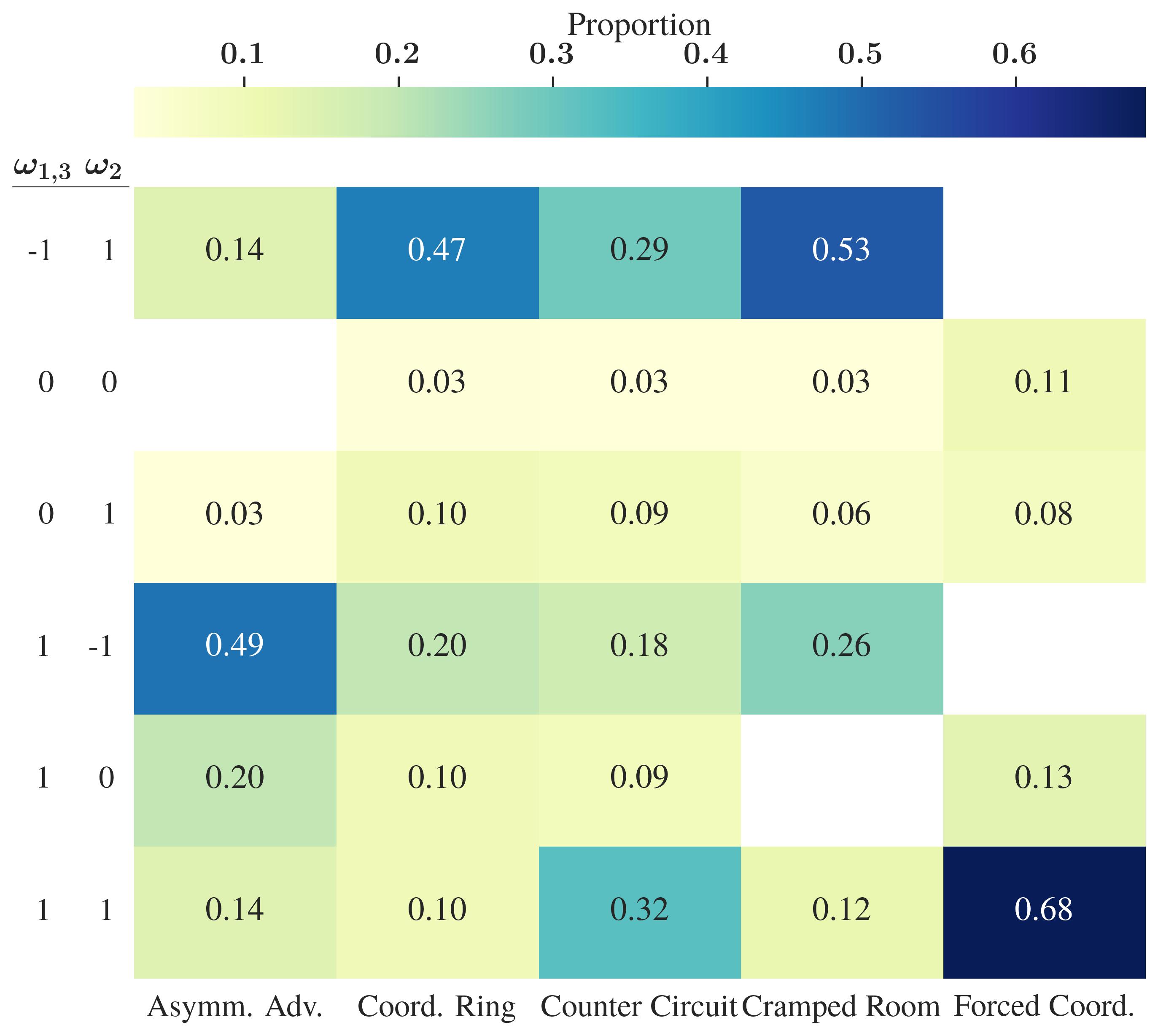}
        \caption{Selection proportion of each $\omega$ combination.}
        \label{fig:weight_prop_heatmap}

    \end{subfigure}%

    \caption{Score and weight heatmaps. We look at the expected value of selecting a particular combination of $\omega$ values in the choice phase of the experiment and every combination of $\omega$ values. Blank cells represent those that were not selected in the choice round.}
    
    \label{fig:score_weight_heatmaps}

\end{figure*}


While Experiment 1 validated the effectiveness of BS as a policy in human-AI collaboration, Experiment 2 exposes the $\omega$ controls to participants in order to measure their preferences for AI behaviors in shared workspace collaboration. 

\paragraph{Design.} We recruited 116 participants (45.5\% female, 2.8\% other; avg. age $40.90\pm 10.83$) through Amazon Mechanical Turk. Participants received a base payment and an incentive bonus for each dish delivered. Each participant interacted with three variations of the same BS policy:

\begin{enumerate}
    \item Controllable: Participants adjusted $\omega$ settings for two behaviors: ``Delivering Dishes'' (encompassing the delivery act and plating rewards) and ``Onions in Pot.'' They could be set to Encourage, Discourage, or Neutral. These settings were displayed throughout the round.
    \item Fixed: $\omega$ settings were randomly determined for each participant and displayed but could not be altered. These settings remained constant across rounds for each participant. We did not sample weights that corresponded to discouraging both behaviors to avoid degeneracy.
    \item Hidden: No behavioral settings were displayed, and participants were informed that the partner’s behavior was unspecified. The settings were randomized but fixed for each participant. Again, we did not sample weights that corresponded to discouraging both behaviors.
\end{enumerate}


After reading instructions and completing tutorials on game mechanics, AI interaction, and control usage, participants were assigned to two of five possible layouts. For each layout, they played ten rounds: three for each condition (Controllable, Fixed, and Hidden) and a final round with their choice of partner.
 
Before interaction with the controllable partner, participants specified the desired behaviors on a three-point scale: Discourage ($\omega=-1$), Neutral ($\omega=0$), and Encourage ($\omega=1$) for each behavior. The settings were then constant throughout the round and were displayed to the participants. Similar presentation was used in the Fixed rounds and this information was omitted entirely in the Hidden rounds. After completing three of each, participants selected their choice for their tenth and final round on that layout. 
 
 After each round, participants completed a brief survey indicating their feedback on a continuous scale, which we discretized into 21 buckets. Participants observed ``Strongly Disagree'' on one end and ``Strongly Agree'' on the other. In each type of round, they were instructed to indicate the extent to which they agreed with the statements: (1) ``my partner was enjoyable to work with,'' (2) ``my partner's behavior was predictable,'' and (3) ``my partner was effective as a teammate.'' In the Controllable and Fixed rounds, the statement ``my partner followed its behavior settings'' was also included. 


\paragraph{Hypotheses.} In Experiment 2, our hypotheses relate to preferences for control and how participants use behavioral controls. We hypothesize (H2.1) that the participants will show a significant subjective preference for the controllable partner over the fixed partner, keeping the objective performance constant. However, our second hypothesis (H2.2) is that this preference is moderated by the deviations the partner makes from the specified controls. Subjective preferences will be measured by responses to the post-round survey and the observed choices in partner selection. Lastly, motivated by existing literature on predictability preferences, we hypothesize (H2.3) that participants will use controls to elicit more predictable behavior from their partners. 


\paragraph{Results}

We first test H2.1 by examining the survey responses from each round. The results of mixed-effects regression models for each response are shown in Table~\ref{tab:reg_results}, columns (1)-(3). We find that, all else equal, participants have a strong subjective preference for controllable partners over hidden-setting partners in terms of their evaluated predictability, effectiveness, and enjoyability as a partner. Indeed, when participants are able to control their partner's behavior, the partner is judged to be more predictable, more effective, and more enjoyable ($p
<0.001$). These results support H2.1 alongside the choice results in Figure~\ref{fig:choice_probs}. All else equal, participants are most likely to select interactions with the controllable partner than either the fixed or hidden-setting partners. 

In evaluating H2.2, we run additional mixed-effects regressions for only partners with observable $\omega$ settings (Fixed and Controllable). We incorporate the response of whether the participants judged their partner to have followed the set behavior settings (``Followed Settings'') as well as the interaction with the condition. 

The results are shown in Table~\ref{tab:reg_results}, columns (4)-(6). There are no significant differences in perceived predictability between the two types of partners. However, the main effects persist, with the controllable partner being evaluated as more effective and enjoyable compared to the fixed-$\omega$ partners. Participants evaluate policies as more predictable, effective, and enjoyable when they adhered more closely to their behavioral settings ($p<0.001$). We find support for H2.2, as the interaction term between following settings and the controllable condition is significant for effectiveness and enjoyability. Adherence to behavioral settings is accentuated in its effects when participants had control over the settings, providing evidence that subjective preferences for control are moderated by deviations to dictated settings.

Finally, we test H2.3 by examining both the performance by $\omega$ values as well as the choices that participants make over the weights in the final episode of their interactions. We hypothesize that a preference for predictability would manifest itself in weight selection through the choice of disparate values for each behavior. For example, discouraging an agent from delivering dishes and encouraging it to put onions in the pot would make that agent more predictable, as it is biased towards a single behavior. This can be thought of as reducing the entropy over behaviors, whereas setting both weights to 0 or 1 would increase the entropy over behaviors as they are uniformly biased. Expressing a preference for predictability would manifest itself as an encouragement of one behavior ($\omega=+1$) while discouraging the other ($\omega=-1$). %

In Figure~\ref{fig:score_weight_heatmaps}, we show both the average score given a layout and weight combination (Figure~\ref{fig:score_heatmap}) and the proportion of times each weight combination is chosen for each layout (Figure~\ref{fig:weight_prop_heatmap}). If humans were solely utility maximizers, we would expect a correspondence between the highest expected value $\omega$ setting and their choice of $\omega$. On the majority of layouts, we see evidence in support of our hypothesis. Indeed, in Asymmetric Advantages, Coordination Ring, and Cramped Room participants favor encouraging one behavior and discouraging the other despite equivalent or better options in terms of expected value. In the remaining two layouts, Counter Circuit and Forced Coordination, we do not see direct support. In the former, participants are split between two suboptimal weight combinations: $\{\omega_{1,3}=-1, \omega_{2}=1\}$ and $\{\omega_{1,3}=1, \omega_{2}=1\}$. The first aligns with our hypothesis, while the latter does not. In Forced Coordination, we observe the vast majority picking the option that makes intuitive sense for the layout: the AI must accomplish all tasks, so all behaviors should be encouraged. These results provide some indication that participants are valuing subjective preferences for predictability over objective outcomes. 
\section{Discussion}

This work contributes to the growing literature on human-AI teaming by investigating human preferences for controllability and predictability in a shared workspace task. We introduced BS, a reinforcement learning algorithm that facilitates the training of robust policies and allows humans to explicitly control AI behavior. Our first experiment validated the zero-shoot coordination capabilities and robustness of BS. Our second experiment provides empirical evidence for the impact of controllability and predictability on subjective outcomes in human-AI collaboration.

In particular, we demonstrated that human preferences extend beyond objective performance. Users overwhelmingly preferred AI teammates that they could control, keeping objective performance constant. When participants had the ability to specify the behavior of their AI teammates, participants reported higher levels of perceived effectiveness and enjoyment in their partners. This was also reflected in choice behavior, as participants consistently favored controllable partners when given a choice of which they would interact with. 

Our results also underscore the importance of predictability and alignment of AI behavior with human expectations, especially when users have control. Participants evaluated AI teammates less favorably when controllable partners failed to adhere to specified behavior settings, even when their overall task performance was not affected. This effect was accentuated when participants controlled the behavior of their partners: perceived deviations are more costly when the human was the one to specify the settings.

The presence of subjective preferences for control sheds light on how subjective preferences can influence the perceived success of human-AI teams. All else equal, participants valued having control over their teammate's behavior. This has implications for designing agents with a broader definition of human-AI complementarity, beyond objective outcomes. Subjective preferences and human expectations should be taken into account and the trade-offs of objective and subjective performance should be fully considered. These considerations can facilitate more meaningful human-AI interaction and adoption of AI in real-world settings.


\bibliographystyle{named}
\bibliography{ijcai25}



\end{document}